\documentclass[12pt]{article}

\usepackage{amsmath,amsfonts,epsfig,color,latexsym}

\newcommand{\be}{\begin{equation}}
\newcommand{\ee}{\end{equation}}
\newcommand{\bea}{\begin{eqnarray}}
\newcommand{\eea}{\end{eqnarray}}

\let\newsection=\section
\renewcommand{\section}{\setcounter{equation}{0}\newsection}

\begin{document}

\begin{flushright}
hep-th/0702037\\
TIT/HEP-565
\end{flushright}
\vskip.5in

\begin{center}

{\LARGE\bf A "black hole" solution of scalar field theory }
\vskip 1in
\centerline{\Large Horatiu Nastase\footnote{email: nastase@phys.titech.ac.jp}}
\vskip .5in

\end{center}
\centerline{\large Global Edge Institute,}
\centerline{\large Tokyo Institute of Technology}
\centerline{\large Tokyo 152-8550, Japan}

\vskip 1in

\begin{abstract}

{\large A black hole-like solution to a toy model scalar field action for the pion is 
presented. It has a "horizon" that traps pion field information (it cannot escape in finite
time) and is thermal, with a finite and calculable temperature, $T=\sqrt{2}/(\pi\alpha)$, with
$\alpha$ a coupling parameter. The action is a scalar DBI action (D-brane action), coupled 
to a particular fixed source term. The DBI action by itself has "catenoid" solutions that  
have horizons with infinite temperature and trap only high energy information. It is also 
proven that the unique scalar action that admits thermal horizons is of DBI type at leading 
order, making the D-brane action special. The existence of this "pionless hole" solution means
that {\em aparent} information loss is not a feature of gravity theories (via black holes), 
but even a simple scalar theory can exhibit it. This is as it should, since the "pionless
hole" is a toy model for the AdS-CFT dual to a black hole. 
}

\end{abstract}

\newpage

\section{ Introduction}

Ever since the original paper of Hawking \cite{hawking} showing that black holes radiate 
thermally with a temperature dependent on the properties of the horizon (through the 
surface gravity at the horizon, $k$), people have been struggling to understand how this is 
possible. The presence of the horizon that traps information and radiates thermally seems to 
be in contradiction with quantum mechanics: the collapse of a pure quantum mechanical state 
into a black hole and its subsequent radiation as a mixed thermal state violates unitarity. 
It was realized that this is not merely a problem having to do with the unknown quantum gravity 
governing the singularity, but a paradox: the thermal property (and information loss) is associated with the horizon, which has no strong curvatures, so quantum gravity should not be 
needed. Within string theory, the belief is that there should be no information loss. Indeed, 
the entropy of extremal black holes has been counted by realizing them as D-brane states
\cite{sv,cm2}, and near-extremal black holes were shown to radiate thermally as the corresponding D-brane system, with the outgoing information encoded in "greybody factors"
\cite{ms}. One might argue that these resolutions of the paradox of black hole thermality 
involve quantum gravity, but perhaps the essential ingredient is rather a new formalism allowing
for the possibility of thermal radiation (though not blackbody!) from a pure quantum system. 

In this paper I will not attempt to find a solution to the paradox, but rather to sharpen 
the question. I will show that the same properties usually associated with black holes are present already in scalar field theory, and thus the resolution of the paradox cannot be 
just the existence of an unknown quantum gravity. Indeed, one certainly understands quantum 
field theory, and a correct treatment should find no breakdown of unitarity. And yet we 
will find a solution that has a horizon that traps scalar field information (it cannot escape in a finite time) and radiates thermally at a finite temperature just like a black hole. 
This is no coincidence, as the model we use is a toy model for the AdS-CFT dual of a black 
hole. In a series of papers \cite{kn,kn2,nastase2,nastase,nastase3,nastase4} the high energy 
small angle scattering (fixed $t$, $s\rightarrow\infty$) in QCD was analyzed using AdS-CFT
\cite{malda} a la Polchinski-Strassler \cite{ps}, and it was found that black hole production 
dominates in the dual. In particular, it was argued that for the collisions observed at RHIC,
the fireballs produced are dual to (analogs of) black holes living on an IR brane \cite{nastase}
(a related proposal was put forth in \cite{amw} where a black hole was considered to be dual to
a large N gauge theory "plasma ball").
The pion field action dominates the physics and is dual to the action of 
a D-brane, thus the pion field should have an action $S=\int d^4 x \sqrt{1+(\partial_{\mu}
\phi)^2}$, as already used by Heisenberg \cite{heis,kn2} to obtain the saturation of 
the Froissart bound \cite{frois}. 

In \cite{nastase3} it was argued that the "catenoid" solution of the scalar DBI action 
\cite{cm,gibbons,review} (itself similar to the BIon solution of the usual electromagnetic
 Born-Infeld action \cite{bi}) should correspond to the black hole in the 
dual via AdS-CFT. It was however found that even though the solution has a thermal horizon, 
similar to Unruh's hydrodynamics "dumb holes" \cite{unruh}, 
the temperature is infinite, and only the high energy modes are trapped at the horizon 
(see \cite{volovik} for extensions and realizations of the "dumb hole" idea in condensed matter 
systems). Moreover, it was assumed there that the 
nucleons themselves should be solitonic solutions of a correct Born-Infeld-type pion field 
action, Skyrmion-like, so it was assumed there should be no need to write a coupling of the 
pions to a nucleon source for the "catenoid" solution. I will review the pure DBI action 
case in section 2.

In this paper, we will instead start with the DBI pion action coupled to a nucleon source,
$\int \phi \bar{N} N$, with the nucleon source $\bar{N}N$ being replaced by an ad-hoc 
fixed spread-out distribution, $\alpha/ r^2$. In section 3,
I will show that this is enough to make the
temperature finite, and trap all information in the pion field at the horizon of the solution.
This is then a toy model for the dual of the black hole, as the particular source term $\alpha
/r^2$ has no good physical justification other than it works, but one should take it as a 
proof of principle. I will however prove in section 4
that the most general scalar field action that 
has a thermal horizon analog to Unruh's case is near the horizon of DBI type plus corrections,
and the only correction that gives a finite temperature without any extra assumptions about the 
high energy behaviour is a coupling to a source term that decays faster than $1/r$. 
For completeness, I will also sketch in section 5 the steps of Hawking's derivation of the temperature within the context of the DBI action. 

The existence of this solution within scalar field theory means that perhaps one also 
needs a new formalism to deal with the possibility of thermal emission from a pure quantum 
state, in quantum field theory as well as in gravity. And in any case, the information loss 
paradox is as paradoxical in quantum field theory as it is in gravity.

\section{Review of the DBI action case}

This section is a review of parts of \cite{nastase3}. The DBI action (with delta function 
source)
\be
S=\beta^{-2}\int d^4 x [\sqrt{1+\beta^2(\partial_{\mu}\phi)^2}-1]+\int d^4 x \phi (\bar{C}
\delta(r)) 
\label{dbi}
\ee
has the "catenoid" solution \cite{cm,gibbons,review}
\be
\phi(r)=\bar{C}\int_r^{\infty} \frac{dx}{\sqrt{x^4-\beta^2\bar{C}^2}}
\ee
that has an aparent singularity at $r_0=\sqrt{\beta \bar{C}}$, where $\phi '(r)\rightarrow
\infty$, but $\phi (r)$ is finite. This aparent singularity acts
as the horizon of a black hole. The above action has a delta function source needed to 
obtain the catenoid solution, however a proper, SU(2)-invariant action for the pion, 
defined in eq.7.2 of \cite{nastase3} does not need a singular source for a solution with 
horizon, and in the presence of apropriate higher order corrections has a topological 
soliton solution representing nucleons, like the skyrmion. In most of the following, we will put
$\beta=1$ for simplicity. Note that $\bar{C}$ is a scalar field charge (at infinity, the 
field is $\phi \sim \bar{C}/r$), quantized in the quantum theory. We also note that, since the
solution never reaches $r=0$, the source term could be argued to be unnecessary, but we kept 
it to point out that the solution is not solitonic in nature.

The above scalar DBI action was found by Heisenberg \cite{heis} to be needed to obtain the 
saturation of the Froissart unitarity bound, $\sigma(s)=(\pi/m_{\pi}^2) \ln^2 s/s_0$. He has
found that in the high energy limit (fixed t, $s\rightarrow\infty$) of hadron scattering, the 
hadrons are irrelevant, and the effective interaction is between pion field shockwaves 
sourced by the hadrons. He needed the above DBI action to saturate the bound. More precisely,
the action had a pion mass term, $m_{\pi}^2\phi^2$ inside the square root (and no source term),
but we will neglect the mass term in the following. 

The fluctuation equation around the catenoid solution, for $\phi=\Phi(r)+\delta \Phi$
($\Phi(r)$ is the catenoid) is
\be
-\partial_t \frac{1}{\sqrt{1+(\vec{\nabla}\Phi)^2}}\partial_t \delta \Phi
+ \partial_i \frac{1}{\sqrt{1+(\vec{\nabla}\Phi)^2}}(\delta^{ij}-
\frac{\partial^i \Phi}{\sqrt{1+(\vec{\nabla}\Phi)^2}}
\frac{\partial^j \Phi}{\sqrt{1+(\vec{\nabla}\Phi)^2}})\partial_j \delta \Phi=0
\label{fluctu}
\ee
The ratio of the coefficients of $\partial_i\delta^{ij}\partial_j \delta\Phi$ and 
$\partial_t^2\delta \Phi$ tends to zero at the horizon, where $\Phi ' (r)\rightarrow
\infty$. This is suggest why we called the aparent singularity a horizon, and also 
suggests a comparison with Unruh's "dumb holes" \cite{unruh} in ultrasonic hydrodynamic fluid flow ("sonic booms"). In that case, the hydrodynamic flow with speed v is irrotational 
($\vec{\nabla} \times \vec{v}=0$), so is described by a potential $\Phi$ by $\vec{v}=\vec{\nabla
}\Phi$, with fluctuation equation
\be
\frac{1}{\rho}(\frac{d}{dt}+\vec{v}\cdot\vec{\nabla}+(\vec{\nabla}\cdot\vec{v}
))\frac{\rho}{c^2}(\frac{d}{dt}+\vec{v}\cdot\vec{\nabla})\delta\Phi-\frac{1}{\rho}
\vec{\nabla}( \rho \vec{\nabla}\delta\Phi)=0
\ee
which is exactly the equation of motion for a scalar field in a 
curved spacetime, $\partial_{\mu}
\sqrt{g}g^{\mu\nu}\partial_{\nu} \delta\Phi=0$ if the metric is given by 
\be
\sqrt{g}g^{\mu\nu}=\rho\begin{pmatrix} \frac{1}{c^{2}} & \frac{v^i}{c^2}\\
\frac{v^j}{c^2} & \frac{v^i v^j}{c^2}-\delta ^{ij}\end{pmatrix}
\label{metricmap}
\ee

After finding $g_{\mu\nu}$ in 4d and defining a new time 
coordinate by $d\tau=dt +v^i dx^i/(c^2-v^2)$, we get the line element
\be
ds^2= \frac{\rho}{c}
[ (c^2-v^2)d\tau^2-(\delta^{ij}+\frac{v^i v^j}{c^2-v^2})dx^i dx^j]
\label{linele}
\ee
or, in the case of a radial flow
\be
ds^2=\frac{\rho}{c}
 [ (1-v^2/c^2)c^2d\tau^2-\frac{dr^2}{1-v^2/c^2 }-r^2 d\Omega^2]
\label{radflow}
\ee
In the new coordinates, the scalar wave equation is 
\be
\partial_0 \frac{\rho/c^2}{1-v^2/c^2}\partial_0\delta\Phi +\partial_i
\rho (\frac{v^i v^j}{c^2}-\delta^{ij})\partial_j \delta\Phi =0
\label{neweq}
\ee
We see that this is the same equation as (\ref{fluctu}) if we make the identifications
\be
c^2=1+(\vec{\nabla}\Phi)^2, \;\;\; \rho=\frac{1}
{\sqrt{1+(\vec{\nabla}\Phi)^2}};\;\;\;  v^i=\partial^i \Phi
\label{velo}
\ee
where the last identification means that indeed $\Phi$ acts as the fluid potential. 

As Unruh described, for the existence of a thermal horizon, all we need is that the 
propagation of the scalar fluctuation $\delta\Phi$ obeys the wave equation in a 
black hole metric, as found above (both for the "dumb holes" and for the catenoid), and 
that as usual, one has a nonzero surface gravity at the horizon for that metric. Indeed, 
that was the only assumption in the calculation of Hawking of the black hole temperature
\cite{hawking}. For completeness we will sketch a few of the steps of that calculation 
in section 5, but we can now just take Hawking's result for the temperature, $T= k/(2\pi)$, 
where $k$ is the surface gravity at the horizon.

For a static, spherically
symmetric solution with only $g_{rr}(r)$ and $g_{tt}(r)$ nontrivial
(and possibly an r-dependent conformal factor for the sphere metric), one can 
easily calculate that 
\be
(2k)^2= \lim_{horizon} \frac{g^{rr}}{g_{tt}}(\partial_r g_{tt})^2
\ee
For a Schwarzschild black hole, $g^{rr}=g_{tt}$ and 
we calculate that $k=1/(4MG)$, as known. For the ``dumb hole'', using 
the above map to a curved spacetime (\ref{metricmap}) we get that 
\cite{unruh}
\be
(2k)^2 = \{\frac{1}{\rho}\partial_r[\rho c (1-v^2/c^2)]\}^2 |_{v=c}
\Rightarrow T=\frac{1}{4\pi}\frac{1}{\rho}\partial_r[\rho c (1-v^2/c^2)] 
|_{v=c}
\label{tempera}
\ee
In Unruh's case, where $\rho$ and $c$ are nonzero and finite
at the horizon, 
one gets $T=(dv/dr|_{v=c})/(2\pi)$, but in our case that is not true.

For the fluctuation around the catenoid, one gets
\be
2k=|\sqrt{1+\Phi '^2}\frac{d}{dr}[\frac{1}{1+\Phi'^2}]|_{r=r_0}
\label{sfgrav}
\ee

and putting the explicit form of $\Phi (r)$, we get 
\be
T= \sqrt{\frac{r_0}{4(r-r_0)}}
\frac{1}{\pi r_0}
\ee

thus the temperature is infinite.

Therefore the horizon is thermal (though of infinite temperature in this case), but another 
property that makes the black hole horizon special (and is related to its aparent lack 
of unitarity) is the trapping of information. All information cannot come out of the 
horizon: it takes an infinite time for a geodesic (massless particle) to come out of it due
to the infinite gravitational redshift at the horizon.

The scalar field theory is in flat space, so obviously some information can travel to and from 
the horizon in finite time. But the essential point is only what happens to {\em scalar 
field information}. And we know that light in a medium for instance can be slowed down even to 
observable speeds. So we have to see what happens to the propagation of information, defined 
by the {\em characteristic surface}, or in other words to the phase and group velocities of 
propagation, $c_{ph}=\omega /k$ and $c_{gr}=d\omega/dk$. And because the equation of motion 
of scalar field fluctuation is the same as the one in a black hole background, the characteristic surface and phase and group velocities are the same as for motion in a black 
hole background. The time delay at the horizon is defined by $ds^2=0$ in the equivalent 
black hole background, so by (in hydrodynamics variables)
\be
\int dt=\int^{horizon}\frac{dr}{c(1-v^2/c^2)}
\ee
and an infinite time delay at the horizon means $d(c(1-v^2/c^2))/dr$ finite at the horizon, 
which is almost the same as the condition for finite temperature found above, that 
$d(\rho c(1-v^2/c^2))/(\rho dr)$ is finite at the horizon. For the catenoid solution, neither 
is true, so we have an infinite temperature and a finite time delay. However, calculating 
the phase and group velocities one finds
\bea
&&c^2_{ph}=\frac{\omega^2}{k^2}=\frac{r^4-r_0^4}{r^4}+\frac{6i}{k}\frac{r_0^4}{
r^5}\equiv a+\frac{b}{k}
\rightarrow \frac{6}{k r_0} \; \; {\rm as\; r\rightarrow r_0}
\nonumber\\&&
c_{gr}= \frac{d\omega }{dk}= \frac{a+b/(2k)}{a+b/k}\rightarrow \frac{1}{2}
\sqrt{\frac{b}{k}}\rightarrow \frac{1}{2} \sqrt{\frac{6}{kr_0}}
 \; \; {\rm as\; r\rightarrow r_0}
\eea
so at least
high energy modes do get an infinite time delay in the limit. Moreover, one easily sees 
that the condition that all modes get zero phase and group velocities at the horizon is the 
same as the condition of infinite time delay from the equivalent black hole metric.

\section{Toy model}

In this section we will describe a toy model that has both infinite time delay for information
and finite temperature. We will use a spread out (hadron) source for the pion DBI action, of the type $\int d^4 x \phi (\alpha/r^2)$ instead of the delta function source used above. This 
is an ad-hoc procedure, thus we will get a toy model, but it describes well the presence of the 
hadron sourcing the pion field. 

Thus we start with the action 
\be
S=\beta^{-2}\int d^4 x [\sqrt{1+\beta^2(\partial_{\mu}\phi)^2}-1]+\int d^4 x \phi
(\alpha/r^2)
\ee
and set $\beta=1$ as above for simplicity. The equation of motion for a radial solution is 
\be
\frac{d}{dr}(\frac{r^2\phi'}{\sqrt{1+\phi '^2}})=\alpha
\ee
with the solution
\be
\phi (r)=\int_r^{\infty}dx\frac{\bar{C}+\alpha x}{\sqrt{x^4-(\bar{C}+\alpha x)}}
\ee
The position of the horizon, where $\phi '\rightarrow \infty$, is where the function 
$f(r)=r^4-(\bar{C}+\alpha r)^2$ reaches a zero. For small enough $\alpha$ (with respect to 
$\sqrt{\bar{C}})$, the function has a single zero, that is just a perturbation of the $\alpha
=0$ result $r_0=\sqrt{\bar{C}}$. For negative and large enough $\alpha$, there are 2 more zeroes, larger than the first. Thus in between, we have a case where there are only 2 zeroes,
the smaller one being a perturbation of the $\alpha=0$ case, and the larger one is a 
local minimum of f(r), thus locally $f(r)\simeq f_0(r-r_0)^2$. It will be clear a posteriori
why we are interested in this solution. 

Cancelling the constant and linear terms in $r-r_0$, we find this solution to be 
$\bar{C}=-\alpha^2/4, r_0=\alpha/2$ for $\alpha >0$ or $\bar{C}=\alpha^2/4, r_0=-\alpha/2$
if $\alpha<0$. We will consider the case $\alpha>0$ in the following. Then near $r=r_0$, we 
have $\phi '\simeq r_0/(\sqrt{2}(r-r_0))$, or
\be
\phi \simeq (\phi_0+)\frac{\alpha}{2\sqrt{2}}\ln (r-r_0)
\ee
where the constant term will become subleading in the $r\rightarrow r_0$ limit. 

Now we can repeat the calculation of the previous section, and the fluctuation equation in 
this background will be unmodified (since the $\alpha$ term is a source, linear in $\phi$). 
Then the fluctuation equation (\ref{fluctu}), as well as the identifications in (\ref{velo}) 
remain the same. The expression for the surface gravity at the horizon in the equivalent 
black hole metric in term of $\Phi(r)$ remains the same as well, as in (\ref{sfgrav}), just that
now the function $\Phi(r)$ is different. Plugging in $\Phi(r)$ near the horizon, we get 
\be
T=\frac{\sqrt{2}}{\pi\alpha}
\ee
so the temperature is now finite! We also see why we wanted $f(r)\simeq f_0(r-r_0)^2$ near the
horizon, since it gives $\Phi \propto \ln (r-r_0)$, which is the only $\Phi(r)$ that makes 
the surface gravity in (\ref{sfgrav}) finite. Note that if we restore the dependence on 
$\beta$, $\phi$ near the horizon remains the same, but $T$ gets multiplied by $1/\beta$.

We now also have an infinite time delay for scalar field information coming to or from the 
horizon, since now
\be
\frac{d}{dr}[c(1-\frac{v^2}{c^2})]_{r=r_0}= \frac{d}{dr}(\frac{1}{\sqrt{1+\Phi'^2}})|_{r=r_0}
\simeq \frac{2\sqrt{2}}{\alpha}
\ee
and as a result the time delay at the horizon goes like
\be
t\sim \frac{\alpha}{2\sqrt{2}} \ln (r-r_0)
\ee
thus is infinite. 

In order to obtain the phase and group velocities, we substitute
$\delta \Phi= A \; exp(i(\omega t -kr))$ 
(spherical waves) in the perturbation equation (\ref{fluctu}), obtaining near the horizon
\be
\omega^2 \simeq \frac{1}{1+\Phi'^2}(k^2-3ik \frac{\Phi' \Phi ''}{1+\Phi '^2})
\simeq \frac{8(r-r_0)}{\alpha^2}[(r-r_0)k^2+3ik]
\ee
and then we get 
\be
c^2_{ph}=\frac{\omega^2}{k^2}\simeq \frac{8(r-r_0)}{\alpha^2}[r-r_0+\frac{3i}{k}]\rightarrow 0
;\;\;\;\; c_{gr}=\frac{d\omega}{dk}\simeq 
\frac{\sqrt{i}}{\alpha}\sqrt{\frac{6(r-r_0)}{k}}\rightarrow 0
\ee
so both go to zero at the horizon for all k. 

Since this solution acts like a black hole with respect to perturbations in the pion field, 
we will call it a "pionless hole".

\section{Uniqueness of the scalar theory}

Let us now try to understand the generality of the above solution, with thermal horizons and
information trapping, within scalar field theory.

Consider a relativistic lagrangean that depends both on $\phi$ and on its derivatives, through
the combination $(\partial_{\mu}\phi)^2$, i.e. ${\cal L}(\phi, (\partial_{\mu}\phi)^2)
={\cal L}(\phi, -\dot{\phi}^2+(\vec{\nabla}\phi)^2)$. Also consider a static, spherically 
symmetric solution $\Phi_0(r)$. The fluctuation equation in the background of this solution 
will be 
\be
\{   \frac{\delta^2{\cal L}}{\delta \dot{\Phi}^2}(\frac{d^2}{dt^2}\delta\Phi)+\nabla_j
[\frac{\delta^2{\cal L}}{\delta \nabla_i\Phi \delta \nabla_j\Phi}]\nabla_i \delta\Phi
-\frac{\delta^2{\cal L}}{\delta\Phi^2}\delta\Phi\}|_{\Phi=\Phi_0(r)}=0
\label{general}
\ee

We see that again, except for the last, $\Phi$-dependent term, the fluctuation equation 
looks the same as the hydrodynamic one of Unruh (\ref{neweq})!

Thus the condition for the existence of a horizon that traps information and maps to Unruh's 
analysis as above is that the ratio of the coefficients of $\nabla_i\delta^{ij}\nabla_i\delta
\Phi$ and $d^2\delta\Phi/dt^2$ goes to zero at the horizon. This is then
\be
\frac{\delta^2{\cal L}}{\delta \Phi'^2}/\frac{\delta^2 {\cal L}}{\delta\dot{\Phi}^2}|_{\Phi=
\Phi_0(r)}\rightarrow 0
\ee
at the horizon or, for a relativistic lagrangean,
\be
\frac{\delta^2{\cal L}}{\delta ((\partial_{\mu}\phi)^2)^2}/\frac{\delta {\cal L}}{
\delta (\partial_{\mu}\phi)^2}|_{\Phi=\Phi_0(r)}\rightarrow -\frac{1}{2\Phi'^2}|_{\Phi=\Phi_0
(r)}
\ee
at the horizon. If at the horizon $\Phi '_0(r)$ goes to infinity and dominates over $\Phi(r)$
(which seems to be the only way to satisfy the horizon condition, but it seems hard to prove)
we can approximate the lagrangean on the solution near the horizon as 
${\cal L}\sim ((\partial_{\mu}\phi)^2)^n$ and substituting in the above condition we get 
that n=1/2, thus the lagrangean on the solution, looks to leading order near the horizon 
like the DBI one! So the DBI action can only be corrected by terms that are subleading 
near the horizon, in order to have a horizon. 

The identification that makes maps the fluctuation equation (\ref{general}) to the one for 
the hydrodynamic potential in (\ref{neweq}) is 
\bea
&& \rho=2\frac{\delta {\cal L}}{\delta (\partial_{\mu}\phi)^2}\nonumber\\
&& c=\frac{1}{\sqrt{1+2(\nabla\phi)^2(\frac{\delta^2{\cal L}}{\delta((\partial_{\mu}\phi)^2)^2}
/\frac{\delta {\cal L}}{\delta (\partial_{\mu}\phi)^2})}}\nonumber\\
&& v^i=\nabla^i \Phi \frac{\sqrt{-2(\frac{\delta^2{\cal L}}{\delta((\partial_{\mu}\phi)^2)^2}
/\frac{\delta {\cal L}}{\delta (\partial_{\mu}\phi)^2})}}{\sqrt{1+2(\nabla\phi)^2(\frac{\delta^2{\cal L}}{\delta((\partial_{\mu}\phi)^2)^2}
/\frac{\delta {\cal L}}{\delta (\partial_{\mu}\phi)^2})}}
\label{newident}
\eea
 
Of course, the fluctuation equation in (\ref{general}) has an extra term (as already 
noted), which looks like a mass term for the scalar fluctuation in the equivalent black hole 
metric, with mass
\be
m^2=\frac{\delta^2{\cal L}}{\delta \Phi^2}
\ee
which however is not constant in the background of a general solution to a general lagrangean.

Applying the calculation of the temperature of the equivalent black hole metric in 
(\ref{tempera}) to the identification in (\ref{newident}) we obtain the horizon temperature
\be
T=\frac{1}{4\pi}\frac{1}{2\delta {\cal L}/\delta(\partial_{\mu}\phi)^2}\frac{d}{dr}\left[
2\delta {\cal L}/\delta(\partial_{\mu}\phi)^2
\sqrt{1+2(\nabla\phi)^2(\frac{\delta^2{\cal L}}{\delta((\partial_{\mu}\phi)^2)^2}
/\frac{\delta {\cal L}}{\delta (\partial_{\mu}\phi)^2})}\right]_{\Phi=\Phi_0(r)}
\ee

But we have already seen that near the horizon the leading piece of the lagrangean is 
$\sqrt{(\partial_{\mu}\phi)^2}$. We will only analyze subleading pieces of the lagrangean
that depend on $\phi$ only, it seems hard to figure out how to obtain derivative corrections
that are subleading to the DBI lagrangean. Thus we will analyze corrections of the type
a)$\sqrt{(\partial_{\mu}\phi)^2+f(\phi)}$, b) $f(\phi)\sqrt{(\partial_{\mu}\phi)+1}$ and 
c) $\sqrt{(\partial_{\mu}\phi)^2+1}+f(\phi)$, and we will take $f(\phi)=\phi^n, e^{a\phi},
\log \phi$. 

a) ${\cal L}\sim \sqrt{(\partial_{\mu}\phi)^2+f(\phi)}$. 

If $f(\phi)\sim \alpha\phi^n$, we look for a solution that behaves near the horizon like $\phi\sim \phi_0+a(r-r_0)^{-\epsilon}$, and obtain from the equations of motion that 
$\epsilon=1/(n-2)$ and $a=-(2/(\alpha r_0(n-2)))^{1/(n-2)}$. We obtain then for the temperature
\be
T\propto \phi '\frac{d}{dr}\frac{\sqrt{f(\phi)}}{{\phi'} ^2}\propto \frac{1}{\sqrt{r-r_0}}
\ee
which is infinite.

If $f(\phi)\sim \alpha e^{M\phi}$, we see that we can understand it as a limit of large n 
of the previous case, so we try a solution that behaves near the horizon as 
$\phi\sim \phi_0+a\ln (r-r_0)$ (note that we cannot try a power law solution, because it 
will then either dominate $(\partial_{\mu}\phi)^2$ in ${\cal L}$, or be completely negligible,
depending on its sign). Substituting in the equations of motion however, we obtain a contradiction, so there is no such solution (or rather, no $r_0\neq 0$ that acts like a horizon
exists). 

If $f(\phi)\sim \alpha \log\phi$,  near the horizon we obtain the equation $
2\phi {\phi '} ^3/r_0+\alpha \phi \log\phi\phi '' =\alpha {\phi'}^2$, 
which has as a solution $\phi\propto \sqrt{r-r_0}$ (leading behaviour) which means that 
\be
T\propto \phi '\frac{d}{dr}\frac{\sqrt{f(\phi)}}{{\phi'} ^2}\propto \frac{1}{\sqrt{r-r_0}}
\ee
which is infinite. Note that this means $\phi_0=0$; if
not, we obtain a contradiction.

b) ${\cal L}\sim \sqrt{(\partial_{\mu}\phi)^2+1}f(\phi)$.

If $f(\phi)=\phi^n$, near the horizon we get the equation $ \phi ''+2{\phi '}^3/r_0=n{\phi '}
^2/\phi$, with the solution $\phi\simeq \phi_0+a\sqrt{r-r_0}$ as for the pure DBI case. Then 
the temperature is 
\be
T\simeq \frac{\phi'}{f(\phi )} \frac{d}{dr} \frac{f(\phi)}{{\phi '}^2}\sim \frac{1}{\sqrt{r-r_0}}\rightarrow 
\infty
\ee
so we have again infinite temperature. 

If $f(\phi)=e^{M\phi}$, near the horizon we get again $r\phi ''+2{\phi '}^3=0$, with the solution
$\phi\simeq \phi_0+a\sqrt{r-r_0}$ as for the pure DBI case, and the temperature is again 
infinite as for the previous case. 

If we have $f(\phi)=\log\phi$, we get $\phi '' +2{\phi '}^3/r_0={\phi '}^2/(\phi \log\phi)$ 
with solution $\phi=\phi_0+a\sqrt{r-r_0}$ and again $T\propto 1/\sqrt{r-r_0}$, thus infinite. 

c) ${\cal L}\sim \sqrt{(\partial_{\mu}\phi)^2+1}+f(\phi)$. The equation of motion is 
\be
\frac{d}{dr}(r^2\frac{\phi '}{\sqrt{1+{\phi '}^2}})=f'(\phi)
\label{fluctuat}
\ee

The fluctuation equation in the background of a solution is the same as for the pure DBI 
case, with $f''(\phi)$ acting as a mass term for the scalar fluctuation. Thus also the 
temperature is again
\be
T\propto \phi ' \frac{d}{dr}\frac{1}{{\phi '}^2}|_{r=r_0}
\ee
so a finite temperature can only be obtained if $\phi '\propto 1/(r-r_0)$, so $\phi\simeq
\phi_0+a\ln (r-r_0)$. 
On such a solution the left hand side of (\ref{fluctuat}) near the horizon is approximately equal to $2r_0$
(constant), so $f'(\phi)$ should also be constant near the horizon. That excludes $e^{M\phi}$,
whereas for $f(\phi)=\log\phi$ or $f(\phi)=\phi^n$ with $n>1$ it is only true for an intermediate regime, where  $\phi_0\gg a\ln (r-r_0)$, but eventually it also becomes excluded.

In conclusion, the only possibility for a scalar field theory having a solution with a horizon with nonzero and finite temperature is the DBI action with a source coupling 
$f(\phi, r)=\phi g(r)$, where $g(r)$ is a spread out source. That satisfies the equation 
of motion to leading order, but in fact can fail at subleading orders. 

The full solution is then
\be
\phi (r)=\int_r^{\infty} \frac{\bar{C}+(\int x^2 g(x)dx)}{\sqrt{x^4-(\bar{C}+(\int x^2 g(x) dx)^2)}}
\ee
and in order to have a horizon near which $\phi '(r)\propto 1/(r-r_0)$ we see that we need 
the function $f(x)\equiv x^4-(\bar{C}+(\int x^2g(x)dx))^2$ to have a local minimum $x_0$ at 
$f(x_0)=0$ for some 
value of $\bar{C}$. It is easy to check that this is only possible if $g(x)$ decays faster than
$1/x$ at infinity. 

Thus the toy model of the previous section is the unique scalar action that has a horizon with
nonzero and finite temperature, the only modification allowed would be to change the shape of 
the spread out source g(r), with the only constraint to be that it needs to decay faster than
1/r at infinity. That is why we have called the model a toy model for the dual of a black hole,
as it was not clear how to select the source g(r), otherwise it is fixed. 

The uniqueness of the DBI action in this context recalls another such property, of the 
electromagnetic DBI action, which is the the unique nonlinear correction of 
the Maxwell theory that is both causal and has only one characteristic 
surface (generically, there are two) \cite{pleb}.

It is also clear why a solution of scalar field theory that looks like a black hole was not 
considered before: there doesn't seem to be another relativistic action that has such a solution!

\section{Sketch of Hawking's proof within scalar theory}

In this paper I have mentioned that once we have the effective metric in which the perturbation
of the scalar field moves, we can just follow Hawking's derivation to the final result
(as noted already by Unruh \cite{unruh}), but 
it is perhaps instructive to elaborate a bit and address some objections one might have. 

Throughout the paper I have used the map to black holes via Unruh's hydrodynamic "dumb
holes." I have found the construction intuitive, especially in light of the fascinating 
identification of the scalar field $\Phi$ with the hydrodynamic potential. But all we 
needed actually to retain was the parametrization used in (\ref{linele}) of a potential 
black hole-type metric, giving the equation for a scalar field propagating in it, 
$\partial_{\mu}\sqrt{g}g^{\mu\nu}\partial_{\nu}\Phi=0$, as in (\ref{neweq}). Here $\rho,
v^i, c$ can be thought as just parameters of the effective metric.

It is important to stress that Hawking's derivation does not use the dynamics of gravity
(Einstein's equation), only its geometry, i.e. the fact that the propagation of a scalar
field occurs within a 4d black hole space, with the usual equation
$\partial_{\mu}\sqrt{g}g^{\mu\nu}\partial_{\nu}\Phi=0$. In fact, the expression of T in terms 
of the surface gravity at the horizon $k$, shows that we could calculate the temperature for 
a scalar field propagating within a solution to gravity with any field equation, or even 
within a background that is not a solution to the gravity field equations. That is the 
analogy we are pursuing here, as the efective line element in (\ref{linele}) is actually 
defined by a solution to a scalar field theory, not by a solution to some gravity theory.

Hawking's derivation relies on the fact that the vacuum for incoming particles (on ${\cal I}^-$)
differs from the vacuum for outgoing particles (on ${\cal I}^+$), and there is a 
Bogoliubov transformation between them. Thus on ${\cal I}^-$ we have 
\be
\phi =\sum_i (f_i a_i +\bar{f}_i a^+_i)
\ee
where $f_i$ are incoming eigenfunctions (on ${\cal I}^-$)
and the vacuum is defined by $a_i|0_->=0$, whereas on ${\cal I}^+$ the expansion is 
\be
\phi= \sum_i (p_ib_i+\bar{p}_ib_i^++q_ic_i+\bar{q}_ic_i^+)
\ee
where $p_i$ are purely outgoing eigenfunctions (with no Cauchy data on the event horizon)
and $q_i$ have no outgoing component (no Cauchy data on ${\cal I}^+$). There is a Bogoliubov
transformation between the 2 expansions, giving
\be
p_i=\sum_i (\alpha_{ij}f_j+\beta_{ij} \bar{f}_j)
\ee
with particle creation on ${\cal I}^+$ in the vacuum state $|0_->$, 
\be
<0_-|b_ib_i^+|0_->=\sum_j|\beta_{ij}|^2
\ee

So in order to calculate the (thermal) particle creation, the outgoing functions $p_i$ are 
continued back to ${\cal I}^-$, and expanded in incoming waves, finding the $\alpha_{ij}$
and $\beta_{ij}$'s and then deriving the particle creation $<0_-|b_i^+b_i|0_->$. 

The rest of the derivation involves this continuation of the outgoing wavefunctions, which 
is done in the region outside the horizon, so no continuation inside the horizon needs to be done. This is good, since in fact there does not seem to be possible to analytically continue 
inside the horizon in a physically meaningful way. This was analyzed in some detail for the catenoid in \cite{nastase3}, and it was found that the only possible continuation was to another
asymptotic region, which can be chosen to be a large, metastable bubble. This continuation is 
somewhat analogous to the Einstein-Rosen bridge for a Schwarzschild black hole, which connects
two asymptotic regions through the black hole horizon (the Schwarzschild throat). Of course, 
the analogy is not perfect, as the Einstein-Rosen bridge is just another way of foliating 
the complete black hole 4d space in 3d slices, and therefore the bridge is not static, but 
gets created and collapses quickly, before it can be traversed. But all we need for the 
Hawking derivation is the 4d effective metric outside the horizon, and of course the gradient
of the metric at the horizon, determined by the surface gravity $k$. 

The expansion of $f_i$ and $p_i$ for the 4d Schwarzschild black hole
behaves asymptotically as 
\bea
&& f_{\omega  lm}\sim \frac{1}{r\sqrt{\omega }}F_{\omega}(r)e^{i\omega v}Y_{lm}(\theta,\phi)
\nonumber\\
&&p_{\omega  lm}\sim \frac{1}{r\sqrt{\omega }}P_{\omega}(r)e^{i\omega u}Y_{lm}(\theta,\phi)
\eea
and a similar formula will apply in our case. The only difference is in the definition of 
the lightcone expansion parameters u and v, which for the 4d Schwarzschild black hole is
\bea
&& v=t+r_*=t+r+2M \log |\frac{r}{2M}-1|\nonumber\\
&&u=t-r_*=t-r-2M \log |\frac{r}{2M}-1|
\eea
where $r_*$ are "tortoise" coordinates that go to infinity at the horizon and measure the 
time delay for a geodesic going to the horizon ($u=$ constant is an incoming geodesic). Thus
for the "pionless hole" we will have 
\bea
&&v=t+r_*\sim t+r+\frac{\alpha}{2\sqrt{2}}\ln (r-r_0)\nonumber\\
&&u=t-r_*\sim t-r-\frac{\alpha}{2\sqrt{2}}\ln (r-r_0)
\eea

In order to estimate the form of the scattering part of $p_{\omega lm}$ on ${\cal I}^-$, near
the event horizon at $v=v_0$, Hawking uses a trick by parallel transporting vectors defined 
on the event horizon and going to an auxiliary space that has also a past event horizon that 
intersects the actual future event horizon. The essential point is the fact that the affine 
parameter $\lambda$ parametrizing the event horizon is written as $\lambda=-Ce^{-k u}$ with 
$k$ the surface gravity at the horizon. Then the phase $e^{i\omega u}$ of $p_{\omega lm}$
is found to turn for the scattered part $p^{(2)}$ into 
\be
exp(-i\frac{\omega}{k}\log \left(\frac{v-v_0}{CD}\right)) \Rightarrow 
p^{(2)}_{\omega}\sim \frac{1}{r\sqrt{\omega}}exp(-i\frac{\omega}{k}\log \left(\frac{v-v_0}{CD}\right))
\label{ptwo}
\ee
on ${\cal I}^-$, where C and D are constants. Note that the only thing that was used was 
the existence of the effective metric for propagation of the scalar field, which posseses
an ${\cal I}^-$ for incoming waves, an ${\cal I}^+$ for outgoing waves, and an event
 horizon with a nonzero and finite surface gravity $k$. 
 
The last steps of the derivation involve calculating the decomposition of $p^{(2)}$ into 
$f_{\omega '}$ and $\bar{f}_{\omega '}$, calculating the coefficients $\alpha_{\omega\omega '}$
and $\beta_{\omega \omega '}$ and obtaining a thermal spectrum with $T=k/(2\pi)$. 

Finally, let me mention another objection that one might have. In Hawking's derivation it was 
essential that one uses the geometric optics approximation (treat the scalar field 
perturbation as a light ray) up to arbitrarily high energies, since close to the horizon, 
the rays (virtual modes)
that scatter at r=0 experience an arbitrarily high blue shift. In fact, this is 
sometimes considered to be a possible way out of the information paradox (how quantum 
gravity becomes relevant after all). But note that the propagation of virtual modes has 
little reason to be influenced by high energy corrections. Unruh noted that in his case, 
the hydrodynamics 
equations are expected to be relevant only up to a small energy scale, so most of the paper
\cite{unruh} was devoted to the analysis of perturbations due to high energy corrections
(to the effective action). Moreover, the "dumb holes" were used to show that high energy 
perturbations are irrelevant to Hawking's derivation.

In our case, we can treat the relativistic toy model as a good toy model up to arbitrarily 
high energies. The "pionless hole" is a toy model for the AdS-CFT dual of a black hole, 
so the pion action need not be coupled to gravity
(we are in the $M_P\rightarrow \infty$ limit), so from a purely theoretical standpoint 
we could consider it to be a good (effective) action valid up to arbitrarily high energies! 
Also, the DBI action itself
arises as an effective action, with arbitrarily high number of derivatives,
coming in string theory from summing string interactions on a D-brane (with $\beta\sim
\alpha '$ fixed, $g_s\rightarrow 0, M_P\rightarrow \infty$), 
so again it has the right to be consider as a good (effective)
action up to arbitrarily high energies. Yet another argument
for that is the fact that Heisenberg used it for the t fixed, $s\rightarrow\infty$ limit of 
hadron scatering, to obtain the saturation of the Froissart bound. 
Finally, note that in this case (unlike Hawking's case) both the virtual mode and the 
background it moves in have the same origin: the DBI (effective) action, so there is no 
reason to postulate some unknown interaction between the virtual modes and the background.

\section{Conclusions}

In this paper I have presented a scalar field theory model that has a solution, the 
"pionless hole", that has the properties of a black hole: it has a horizon that traps 
all pion field information (it cannot escape in a finite time), and a finite temperature,
$T=\sqrt{2}/(\pi \alpha)$. The model is the DBI action coupled to a fixed, spread 
out source term $g(r)= \alpha/r^2$, and is a toy model for the AdS-CFT dual to a black hole. 

The DBI action itself produces a solution, the catenoid, that has a horizon with infinite 
temperature, and that only traps information in high energy modes. The DBI action arises 
as a good model for the fixed $t$, $s\rightarrow \infty$ behaviour of QCD (Heisenberg showed 
that it is needed to reproduce the saturation of the Froissart bound \cite{heis,kn2}), and 
within AdS-CFT can be used for the dual of a D-brane action, and was indeed expected to have 
an analog of a black hole as solution. 

The particular source term $g(r)$ taken is an ad-hoc model representing the coupling 
to a nucleon (or hadron) source, $\int d^4 x\phi \bar{N}N$, since for the high $s$ 
collisions, the only role of the hadrons is as sources for pion field shockwaves
\cite{heis,kn2}. I have shown that the most general scalar field action that gives a 
solution with a horizon that traps information and has finite temperature is of the 
type considered, except with a general source term $g(r)$ that decays faster than $1/r$
at infinity. One could presumably fix $g(r)$ on physical grounds, though that would have
to be done in the future, and it is clear that it 
needs to decay faster than $1/r$, as that would correspond to a hadron spread out like a 
free scalar field. Since the action considered is the only one giving a black hole-like
solution, it is perhaps not surprising that such a solution has not been thought of before.

The way the "pionless hole" traps information at the horizon is the same as a black hole 
does, namely the phase and group velocities, measuring the propagation of {\em information }
in the nontrivial background, go to zero at the horizon. This is similar to the fact that 
light in a medium will propagate with a smaller velocity (even though we are in Minkowski
space). 

For completeness, I have also sketched Hawking's derivation of the temperature within 
the context of the scalar field solution. I pointed out that only the geometry of gravity 
is needed (not the dynamics, defined by the Einstein equation), which we have in the form 
of the effective metric for propagation of perturbations in the scalar field. Moreover,
we only need this metric outside the horizon, together with the fact that the surface 
gravity at the horizon is nonzero and finite. This is good, since there does not seem to be
any physically meaningful way to analytically continue the solution inside the horizon.
I also pointed out that, unlike Unruh's case \cite{unruh}, we can consider the action to be 
a good (effective) action 
defined up to arbitrarily high energies (at least from a purely theoretical standpoint), 
as it is sometimes argued to be needed in Hawking's derivation. Also, the common (DBI)
origin of virtual modes and background makes irrelevant the postulation of an unknown 
interaction between the two (as is sometimes argued in gravity). 

The implication of this paper for the black hole information paradox is that the resolution
of the paradox cannot be the existence of an unknown theory of quantum gravity, but rather
a new formalism allowing for the formation of an object that radiates thermally (but not 
blackbody!) from a pure quantum state. In other words, since the paradox appears in scalar 
field theory, it is as paradoxical here as in gravity, so its solution should not be 
quantum-gravitational (even if it might involve a theory that includes quantum gravity, 
like string theory). 

In fact, it was argued in \cite{nastase4,nastase3,nastase} that we already have experimental 
evidence for this aparent information loss in field theory: the "fireballs" observed at 
RHIC involve collisions of pure quantum states (2 nuclei, each with about 200 nucleons), 
with subsequent radiation of a mixed state (thermal radiation of tens of thousands of 
the lightest particles in the theory, the pions, together with the rest of the
particles of the theory). In fact, the RHIC collisions were argued to be in the Froissart
regime, in which the pion DBI action is relevant, so the "pionless hole" should in fact 
be a toy model for the fireballs observed at RHIC as well!

{\bf Acknowledgements} 
I would like to thank Aki Hashimoto for discussions.
This research has been done with support from MEXT's program
"Promotion of Environmental Improvement for Independence of Young Researchers"
under the Special Coordination Funds for Promoting Science and Technology.

\end{document}